\documentclass[twocolumn, prl]{revtex4-1}
\usepackage{graphicx}
\usepackage{epsfig}
\begin{document}
\title{Bose-Einstein Condensates in Non-abelian Gauge Fields}
\author{Tin-Lun Ho and Shizhong Zhang}
\affiliation{Department of Physics, The Ohio State University, Columbus, OH 43210}
\date{\today}

\begin{abstract} 
 \end{abstract}

\maketitle

{\bf The recent success of the NIST group \cite{ian1,ian2} in generating abelian gauge field in cold atoms has created opportunities to simulate electronic transports in solids using atomic gases.  Very recently, the NIST group has also announced in a DARPA Meeting the creation of non-abelian gauge fields in a pseudo spin-1/2 Bose gas. While there have been considerable theoretical activities in synthetic gauge fields, non-abelian fields have not been generated until nowÊ
 \cite{Jaksch, Osterloh,Ruseckas,Vaishnav, Satija08, Satija06, Merkl, Goldman1, Goldman2, Goldman3,Wang, Wu, Juzeliunas,Gerbier}. Here, we show that in a non-abelian gauge field, a spinor condensate will develop a spontaneous stripe structure in each spin component, reflecting a ground state made up of two non-orthogonal dressed states with different momenta. Depending on interactions, this ground state can reduce back to a single dressed state. These momentum carrying stripes are the {\em macroscopic} bosonic counterpart of the spin-orbit phenomena in fermions that are being actively studied in electron physics today.}

The beauty of the NIST scheme is its simplicity. It also illustrates the important fact that both abelian and non-abelian fields can be related through a single family of hamiltonian, and suggests  in our opinion a ``generalized adiabatic" scheme for generating gauge fields with increasing complexity. The scheme is the following. Consider the hamiltonian 
$h={\bf p}^2/2M + W({\bf r})$ that operates on an atom with internal degrees of freedom, such as alkali atoms with hyperfine spin $F$.  $W$ is a spatially varying potential in spin space with typical wave-vector $q$. The energy scale for spatial variation of $W$ is then $\epsilon_q= \hbar^2 q^2/2M$.
If $W$ has a group of $L$ states ($L<2F+1$) at the bottom of its spectrum lying within an energy range $\Delta E\ll  \epsilon_q$ and is well separated from all other higher energy spin states by  $\epsilon_q$, then the low energy phenomena of the system can be described within this reduced manifold of $L$ states. By going into a frame in this manifold that transforms away the spatial variations of the spin states, a gauge field emerges. The gauge field is abelian if $L=1$, and non-abelian if $L\geq 2$. Thus, by successively moving the high energy states across  $\epsilon_q$ into the low energy group, one can increase the dimensionality of the low energy manifold and create non-abelian gauge fields with increasingly rich structure. It should be noted that this is very different from the $\Lambda$-scheme in most theoretical proposals, which makes use of a set of dark states sitting distinctly above a short living ground state of the system \cite{Ruseckas}. In contrast, the generalized adiabatic scheme makes use of the lowest energy states. It therefore eliminates the collisional loss and hence intrinsic heating in the $\Lambda$-scheme. 

Before proceeding, it is useful to note the unique features of non-abelian gauge fields.  In the abelian case, a {\em constant} vector potential has no physical effects as it can be gauged away completely. This is not true for the non-abelian case because of its non-commutativity. As a result, a constant vector potential does matter.  Moreover, non-abelian gauge fields inevitably lead to spin-orbit coupling, so any potential (such as a confining trap) that alters particle trajectory also causes spin rotation. This immediately implies significant differences between bosons and fermions. For fermions, Pauli principle forces them into different orbital states. The spin structure of the system is then a result of the contributions of all different occupied states.  In contrast, bosons will search for or even construct (through interaction effects) an optimum spin states, and magnify it to the macroscopic level. The current experiments at NIST have provided us opportunities to study macroscopic spin-orbit effects. 
 
\begin{figure}
\includegraphics[width=0.4 \textwidth]{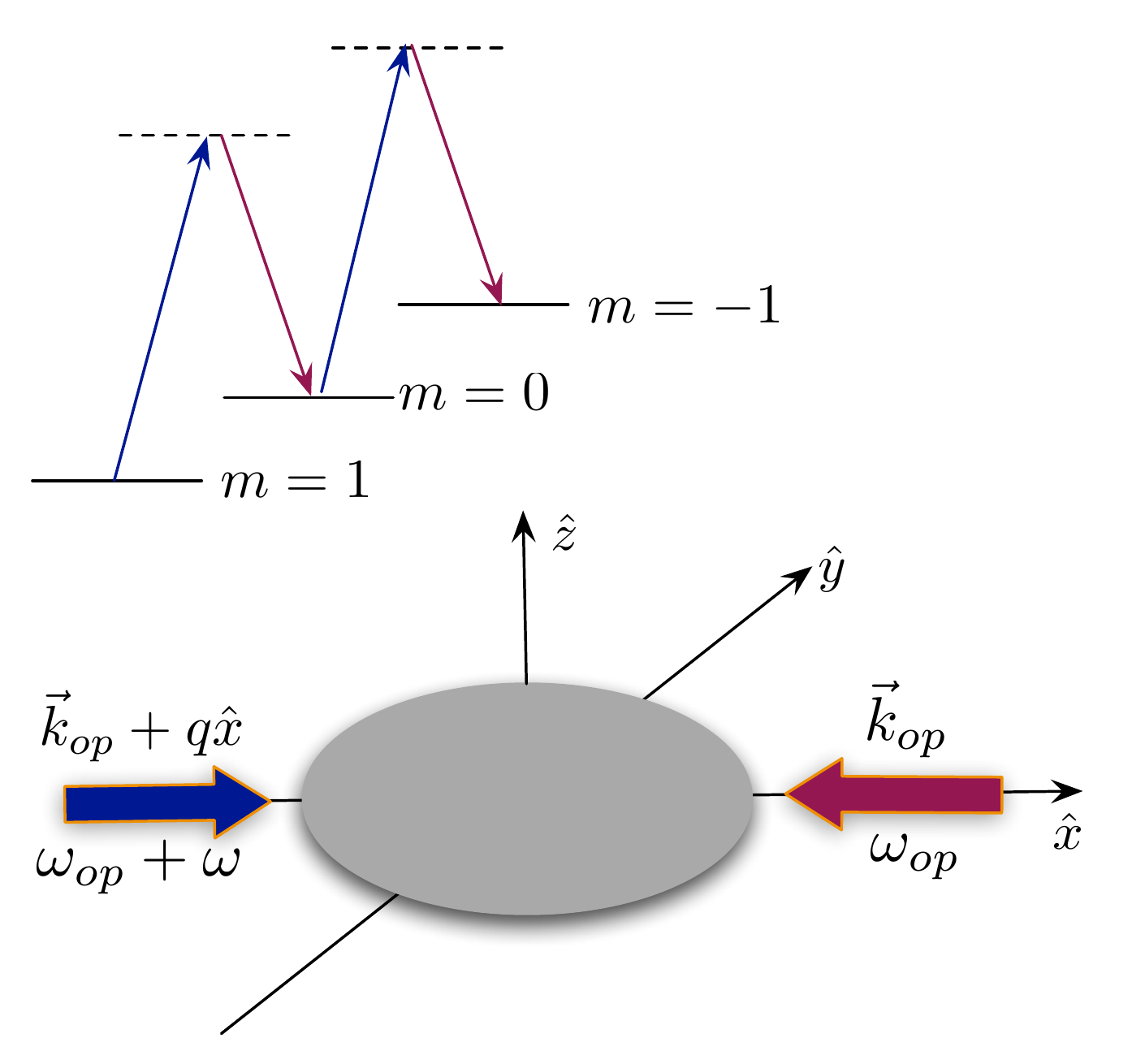}
\caption{Schematics of the experimental setup at NIST. The Raman process consists of two lasers with wave vectors ${\bf k}_{op}+q\hat{x}$ and ${\bf k}_{op}$, frequencies $\omega_{op}+\omega$ and $\omega_{op}$ impinging on the atomic cloud. Atoms excited by the laser will have their momenta increased by $q\hat{x}$ while their  spin projection along $\hat{\bf y}$ is changed by 1, as shown in the energy diagram at top. }
\label{schematic}
\end{figure}

 {\bf (A) The NIST setup and the effective hamiltonian}:
The NIST setup consists of two counter propagating lasers with frequency difference $\omega$ and momentum difference $q$, directed along $\hat{\bf x}$ toward a spin $F=1$ Bose condensate of $^{87}$Rb atoms. 
 There is also a magnetic field directly along $\hat{\bf y}$ with a field gradient. (See Figure \ref{schematic}). The lasers induces a Raman transition in the atom, transferring linear momentum $q\hat{\bf x}$ to the Bose gas while increasing the spin angular momentum by $\hbar$ at the same time. 
 The single particle hamiltonian is $h(t) = \frac{{\bf p}^2}{2M} + W(t)$, where $W(t) = 
 - \hbar\Omega_{y}F_{y} + \hbar \lambda F_{y}^2 -\frac{\hbar \Omega_{R}}{2}[e^{i(qx-\omega t)}(F_{z}+iF_{x}) + h.c. ]$, where ${\bf F}$ is the spin-1 operator. $\hbar \lambda$ is the quadratic Zeeman energy. $\hbar\Omega_{y} = \hbar \Omega_{o} +Gy$ is the Zeeman energy produce by the magnetic field along $\hat{\bf y}$: The $\Omega_{o}$ term is due to the constant magnetic field and the $Gy$ term comes from the field gradient. $\Omega_{R}$ is the
 Rabi frequency in the Raman process. In the frame rotating in spin space along $\hat{y}$ with frequency $\omega$, the hamiltonian becomes static 
$H= h(t=0) +\hbar \omega F_{z}$,  and is given by 
$H={\bf p}^2/2M + W$, 
\begin{equation}
W/\hbar= - \overline{\Omega}_{y}F_{y} + \lambda F_{y}^2 - \Omega_{R}({\rm cos}qx  F_{z}- {\rm sin}qx F_{x}) 
\label{W} \end{equation}
 \begin{equation}
 = e^{-iqxF_{y}} \left( - \overline{\Omega}_{y}F_{y} + \lambda F_{y}^2 - \Omega_{R} F_{z}
 \right)  e^{iqxF_{y}}
\label{W1}  \end{equation}
and $\overline{\Omega}_{y}= \Omega_{o}-\omega +Gy$. 

Eq.(\ref{W1}) shows that $W$ has a very simple level structure in the frame in spin space rotating along $\hat{\bf y}$ with angle $qx$. For simplicity, let us take  $F=1$  and $G=0$.  The following cases are of particular interests:  

\noindent (i) Abelian case: This occurs when $\Omega_{R}\gg \lambda, \epsilon_{q}/\hbar$, with $\omega$ tuned close to $ \Omega_{o}$, hence $\overline{\Omega}_{y}\sim 0$. The ground state in the rotating frame is the $m=+1$ state along $\hat{\bf z}$, isolated from other two states ($m=0,-1$ along $\hat{\bf z}$) by $\sim \hbar \Omega_{R}> \epsilon_q$.

\noindent (ii) Non-abelian case: This occurs when  $\lambda\gg \Omega_{R}, \epsilon_q$, with $\omega$ tuned closed to $\omega= \Omega_{o} - \lambda$.  In this case, the states $m=1$ and $m=0$ along $\hat{\bf y}$ lie at the bottom of the spectrum, separated from the third state $m=-1$ by  $2\hbar \lambda>\epsilon_q$. We shall from now on focus on this case. 


Let  $\hat{\psi}^{\dagger}_{m}$  and $\hat{\phi}^{\dagger}_{m}$ be the operators that create a boson with spin projection $m$ along $\hat{\bf y}$ in the laboratory frame and in the rotating frame in spin space; and $\hat{\psi}_{m} = \left(e^{iqxF_{y}}\hat{\phi}\right)_{m} = e^{iqxm}\hat{\phi}_{m}$. Focusing on the lowest two states $\hat{\phi}_{m}$, $m=1, 0$, the hamiltonian is  
\begin{equation}
\hat{\cal K} = \int \left[  \hat{\phi}^{\dagger}_{m}H_{mn} \hat{\phi}_{n} + \frac{1}{2} \hat{n}_{m}g_{mn}\hat{n}_{n}
-   (V-\mu)\hat{n}\right] 
\label{K} \end{equation}
 where
 $\hat{n}_{m} = \hat{\phi}^{\dagger}_{m}\hat{\phi}^{}_{m}$, $\hat{n} = \sum_{m} \hat{n}_{m}$, $V= \frac{1}{2}M \omega_{T}^2 {\bf r}^2$ is an harmonic trap, $\mu$ is the chemical potential, $g_{mn}$ are interactions between bosons in spin states $m$ and $n$, $g_{10}=g_{01}$, and  
 \begin{equation}
H_{mn} = \frac{\hbar^2}{2M} \left[ \frac{\nabla}{i} + \hat{\bf x} q \left( \begin{array}{cc} 1 &0 \\ 0& 0\end{array} \right) \right]^2 + 
 \hbar \left( \begin{array}{cc} -Gy & \frac{\Omega_{R}}{\sqrt{2}}\\ \frac{\Omega_{R}}{\sqrt{2}} & 0\end{array} \right) .
\label{h}  \end{equation}
When $G=0$, the Schr\"{o}dinger equation
\begin{equation} 
H_{mn}(x)  \chi_{n}(x) = E \chi_{m}(x) 
\label{hSh} \end{equation} 
has the following property. If $\chi$ is  solution of Eq.(\ref{h}), then 
\begin{equation} 
\chi'_{m}(x)  =  e^{i\gamma} e^{-iqx}(\tau_{1})_{mn} \chi^{\ast}_{n}(x), \,\,\,\,\,\, \tau_{1} =  \left( \begin{array}{cc} 0 &1 \\ 1& 0\end{array} \right)
\label{sym} \end{equation}
is also a solution, where $\gamma$ is an arbitrary phase. 



\begin{figure}[h]
\includegraphics[width=0.4 \textwidth]{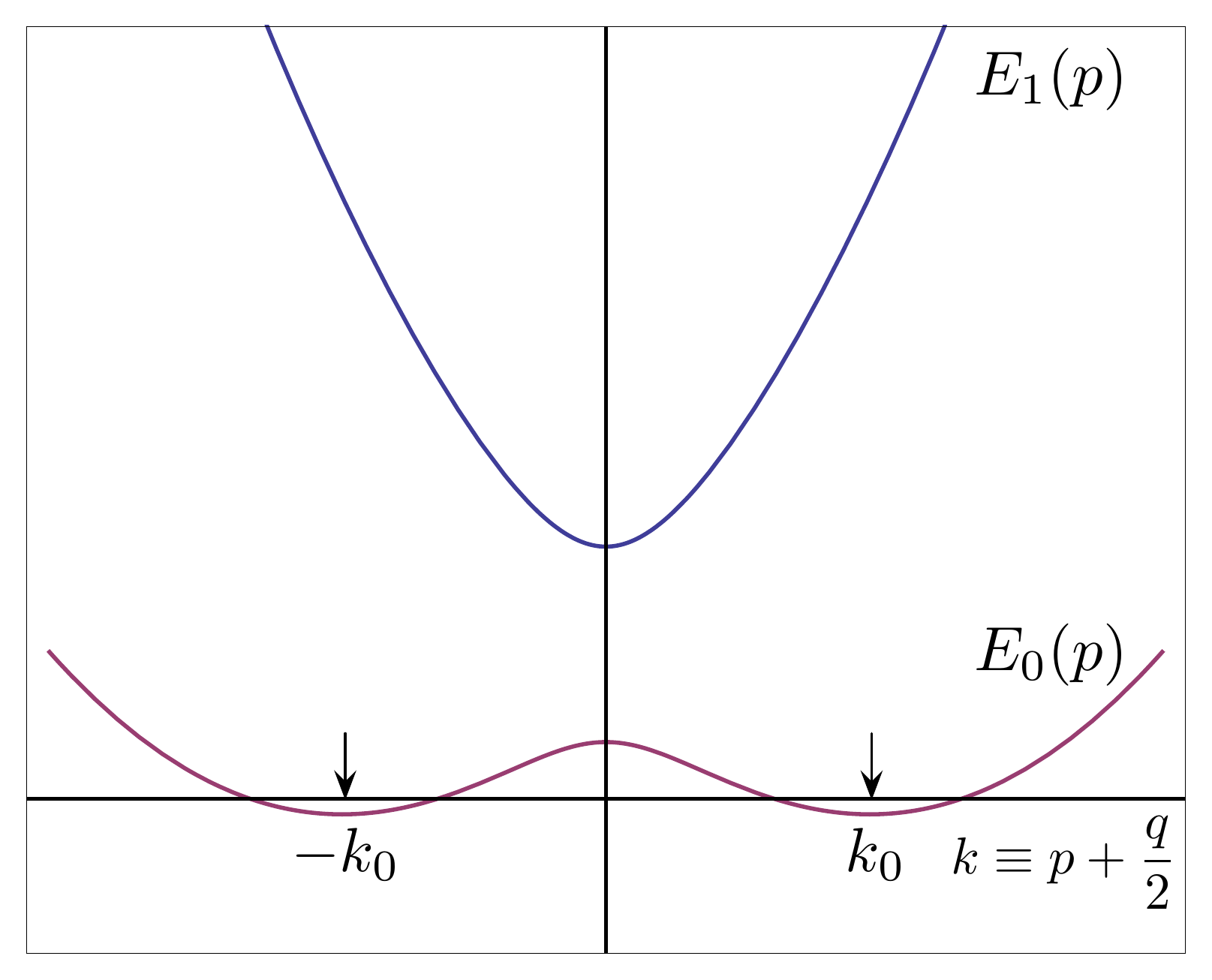}
\caption{The energy levels $E_{0}(p)$ and $E_{1}(p)$ as a function of $k\equiv p+q/2$. The lower branch $E_{0}(p)$ has two degenerate minima  at $k=\pm k_0$, where $k_0=(q/2)\cos\theta$. The energy difference between the lower and upper branch at $\pm k_0$ is $\epsilon_q = \hbar^2 q^2/2m$. }
\label{energy}
\end{figure}

{\bf (B) Single particle ground state}:  In zero field gradient $G=0$, the  momentum eigenstates is of the form 
$\chi^{(p)}_{m}(x) = e^{ipx}\widetilde{\chi}_{m}$, $\widetilde{\chi}\equiv \left(^{u}_{v}\right)$, 
and Eq.(\ref{hSh}) becomes 
\begin{equation}
\frac{\hbar^2}{M} \left(  \frac{k^2 + Q^2}{2} + k
Q\tau_{2} + \ell^2 \tau_{1}\right) \left( \begin{array}{c} u\\v \end{array} \right) = E_{p} \left( \begin{array}{c} u\\v \end{array} \right),
\end{equation}
where we have defined 
\begin{equation}
Q\equiv q/2, \,\,\,\,\,\  k\equiv p+Q.
\end{equation} 
and have 
expressed $\hbar\Omega_{R}$ in terms of a wave-vector $\ell$ and angle $\theta$ for later use, 
\begin{equation}
\frac{\ell^2}{Q^2}  \equiv \frac{ M \Omega_{R}}{\hbar \sqrt{2} Q^2 }= \frac{\sqrt{2}\hbar \Omega_R}{\epsilon_q} \equiv {\rm sin}\theta. 
\label{deftheta}
\end{equation}
The eigenvalues come in two branches, with energies  $
E_{1(0)}(p) = \frac{\hbar^2}{M}\left( \frac{k^2 + Q^2}{2}+(-) \sqrt{(kQ)^2 + \ell^4} \right)$. 
(See Figure \ref{energy}).  The ground states are the minima of $E_{0}(p)$ at
\begin{equation}
p_{\pm}=  \pm k_{o}-q/2, \,\,\,\,\,\,\, k_{o} = +\sqrt{ Q^2 - \ell^4/Q^2}= (q/2){\rm cos}\theta, 
\end{equation}
with energy
\begin{equation}
E_{0}(p_{\pm}) = -\frac{\hbar^2 \ell^4}{2MQ^2}= -\frac{1}{2} \frac{(\hbar \Omega_{R})^2}{\epsilon_q}\equiv E_{o}.
\label{E0} \end{equation}
The energy of the upper branch at these momenta is $E_{1}(p_{\pm})= \epsilon_q - \hbar^2 \ell^4/(2MQ^2)$, which is higher by $\epsilon_q$. It is worth noting that the value of the ground state energy is not of the order $-\hbar\Omega_{R}$, but a higher energy $-(\hbar\Omega_{R})^2/\epsilon_q$.  The wavefunctions at these degenerate ground states are 
 $\chi^{(p_{\pm})}_{m}(x)= e^{ip_{\pm}x} \widetilde{\chi}^{(p_{\pm})}_{m}$, 
  \begin{equation}
\widetilde{\chi}^{(p_{+})}= \left( \begin{array}{c} i {\rm sin}\frac{\theta}{2} \\ {\rm cos}\frac{\theta}{2}\end{array} \right), \,\,\, \widetilde{\chi}^{(p_{-})}=  \left( \begin{array}{c} i {\rm cos}\frac{\theta}{2} \\ {\rm sin}\frac{\theta}{2} \end{array} \right).\end{equation}
Note that the states $\chi^{(p_{\pm})}(x)$ are connected by Eq.(\ref{sym}) with $\gamma=\pi/2$. They are orthogonal due to their different momenta. The spin states, however,  have non-zero overlap, since 
\begin{equation}
\langle p_{+}|p_{-}\rangle =  \widetilde{\chi} ^{ (p_{+}) \dagger} \widetilde{\chi}^{(p_{-})}= {\rm sin}\theta.
\label{overlap} \end{equation}

{\bf  (C) Pseudo spin-1/2 spinor condensate:} 
Condensing in the dressed states $| p^{(\pm)}\rangle$,   the field operator, which has the expansion  $\hat{\phi}_{m}(x) = \sum_{p} \chi^{(p)}_{m}(x) \hat{a}_{p}$, turns into a spinor field of the form 
\begin{equation}
\Phi^{}_{m}(x) = A^{}_{+} \chi^{(p_{+})}_{m}(x) + A^{}_{-} \chi^{(p_{-})}_{m}(x). 
\label{op} \end{equation}
Because of the non-zero overlap Eq.(\ref{overlap}), the density 
 $n_{m}(x) =|\Phi_{m}(x) |^2$ of each spin component will develop a stripe structure.  This can be seen by noting that the total density  $n(x) = n_{1}(x) +n_{0}(x)$ and the ``magnetization" $m(x) = n_{1}(x)- n_{0}(x)$ are given by 
\begin{equation}
n(x)  = |A_{+}|^2 + |A_{-}|^2 + {\rm sin}\theta (A^{\ast}_{+} A_{-} e^{-2ik_{o}x} + c.c.)
\label{nx} \label{n} \end{equation}
\begin{equation}
m(x)  = - {\rm cos}\theta ( |A_{+}|^2 - |A_{-}|^2) .
\end{equation}
Note also that $m(x)$ is independent of $\theta$.  Eq.(\ref{nx}) shows that the contrast of the oscillation is set by the 
overlap ${\rm sin}\theta$, whereas the wavelength of the stripe is $\pi/k_{o}= 2\pi/ (q {\rm cos}\theta)$. Thus, both contrast and wavelength increase with $\theta$ for $\theta <\pi/2$.

The amplitudes $A_{\pm}$ are determined by minimizing the Gross-Pitaevskii functional of Eq.({\ref{K}), which is obtained by replacing $\hat{\phi}_{m}(x)$ with the c-number $\Phi_{m}(x)$, and $\hat{n}_{m}(x)$ with $n_{m}(x)=|\Phi_{m}(x)|^2$. Defining $|{\bf A}|^2= |A_{+}|^2 + |A_{-}|^2$, and $a_{\pm} \equiv A_{\pm}/|{\bf A}|$, the GP functional then reads, 
\begin{equation}
{\cal K} = (E_{o}- \mu)|{\bf A}|^2 + \frac{1}{2} |{\bf A}|^4 G(a_{+}, a_{-}), 
\label{GPfinal} \end{equation}
where  $|{\bf A}|^4 G(a_{+}, a_{-}) = \int g_{mn} n_{m}(x) n_{n}(x)$. Note that while the contributions of $A_{+}$ and $A_{-}$ are separated out in the kinetic energy term due to the different momenta of $\chi^{(p_{+})}(x)$
 and $\chi^{(p_{-})}(x)$, they are coupled through interaction due to the overlap of their spin functions.  For example,  $\int n_{1}^2 (x) = \int [n^2(x) + m^{2}(x) + 2n(x)m(x)]/4$, and the mixing of  $A_{+}$ and $A_{-}$  appears in $\int n^2(x)$.  To minimize ${\cal K}$, we first 
  minimize $G(a_{+}, a_{-})$ with the constraint $|a_{+}|^2 + |a_{-}|^2 =1$ to obtain the optimal value $(a_{+}^{o}, a_{-}^{o})$; and then we have 
 \begin{equation}
|{\bf A}|^2 =( \mu-E_{o})/ G_{o}, \,\,\,\,\,\,\, G_{o} = G(a_{+}^{o}, a_{-}^{o}). 
\label{A2}\end{equation}

\begin{figure}[h]
\includegraphics[width=0.4 \textwidth]{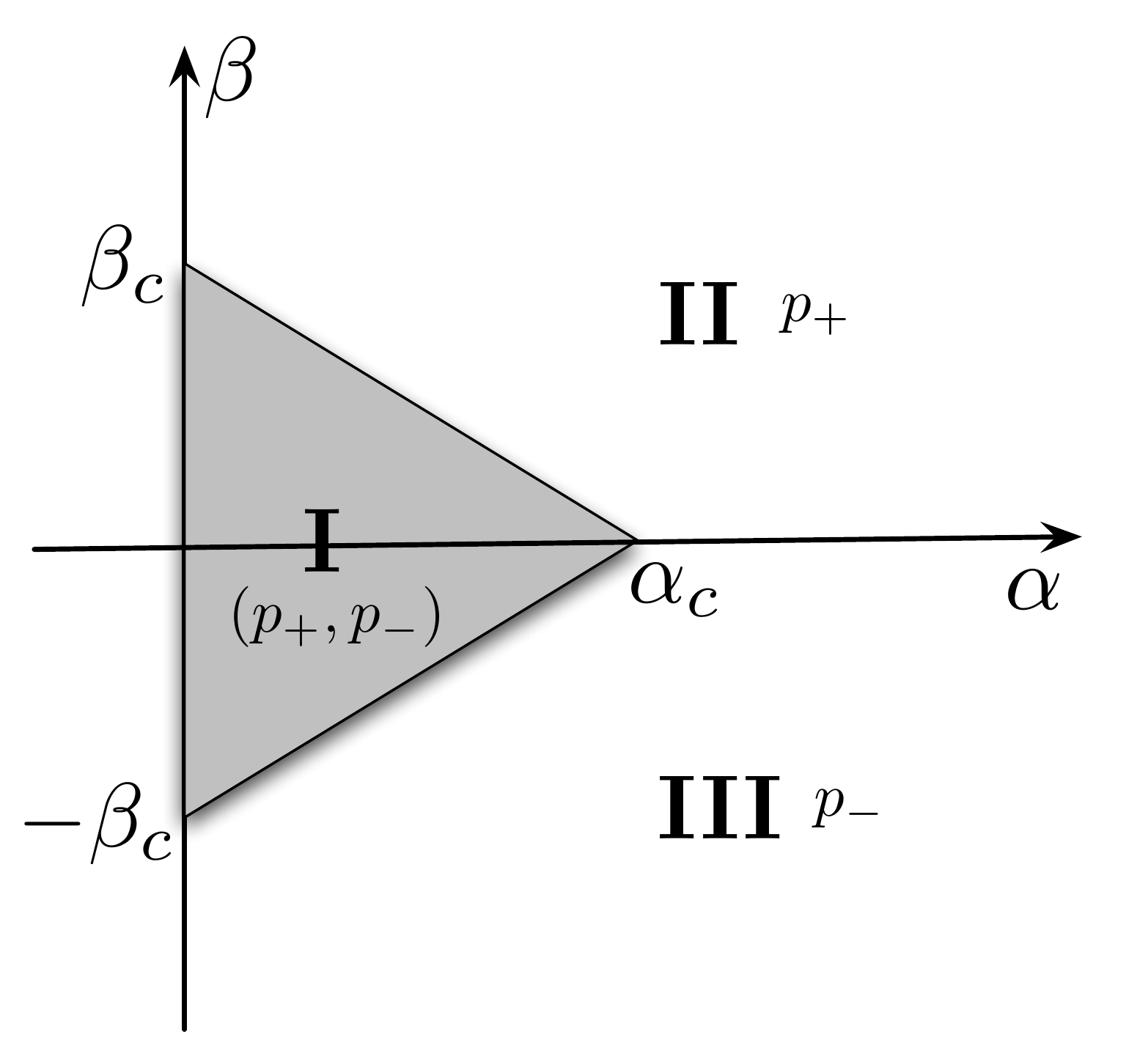}
\caption{The phase diagram of pseudo spin 1/2 Bose gas:  Region {\bf I} is a superposition of two dressed state with momentum $p_{+}$ and $p_{-}$,  {\bf II} and {\bf III} are the single dressed states $p_{+}$ and $p_{-}$ respectively. $\alpha$, $\beta$, $\alpha_{c}$, and $\beta_{c}$ are defined in text. }
\label{phasedigram}
\end{figure}
%


Since the minimization is straightforward, we shall only present the results, which are shown in Figure 3. The phase diagram depends on the  parameters
\begin{equation}
\alpha= g_{10}/g, \,\,\,\,\, \beta= (g_{11}-g_{00})/g, \,\,\,\,\, g=(g_{11}+g_{00})/2. 
\end{equation}
and two numbers $\alpha_{c}$ and $\beta_{c}$ derived from the laser parameter 
${\rm sin}\theta$ defined in Eq.(\ref{deftheta}). They are  $\alpha_c\equiv \frac{2-\tan^2 \theta}{2+\tan^2 \theta}$, and  $\beta_{c} = {\rm cos}\theta (2-{\rm tan}^2\theta)$. 
For $g_{11}, g_{00}, g_{10}>0$, (as in $^{87}$Rb), there are three possibilities: ({\bf I}) Two dressed states, with both $A_{\pm}\neq 0$; 
single dressed state with ({\bf II}) $\chi_{p_{+}}(x)$, ($A_{-}=0$), or   
({\bf III})  $\chi_{p_{-}}(x)$, ($A_{+}=0$).

Phase ({\bf I}) occurs within the triangle shown in Fig.3, 
bounded by the lines $xy_c \pm yx_{c} = x_{c} y_{c}$.  The region exists only when $\alpha_{c}>0$, which means 
  ${\rm sin}\theta < \sqrt{2/3}$. Otherwise, interaction effect will drive the condensate into a single dressed state. 
 In phase $({\bf I})$, the amplitudes are 
\begin{equation}
|a^{o}_{\pm}|^2 = \frac{1}{2}\left(1\pm \frac{\beta/\cos\theta}{2-2\alpha-(1+\alpha)\tan^2 \theta}\right),
\label{apm} \end{equation}
and $G_{o}= G(a^{o}_{+}, a^{o}_{+}) = -\frac{\beta^2}{2(2-2\alpha-(1+\alpha)\tan^2\theta)}+(1+\alpha)(1+\frac{1}{2}\sin^2\theta)$. 
The relative phase between $A_{+}$ and $A_{-}$, however,  remains undermined within the GP approach. This phase can be fixed by  perturbations such as field gradient the breaks the symmetry Eq.(\ref{sym}), or by quantum fluctuation effects that go beyond GP approach. 
As discussed before, the density of each of the spin component $n_{1}$ and $n_{0}$ of this phase has a stripe structure. 
 The case $\beta=0$ ($g_{11}=g_{00}$) is special. In that case, we have 
$|A_{+}|= |A_{-}|$ for $\alpha<\alpha_{c}$. For $\alpha>\alpha_{c}$, the two dressed states $\chi^{(p_{+})}$ and $\chi^{(p{-})}$ are degenerate.

In the presence of a harmonic trap $V({\bf r})= \frac{1}{2}M\omega_{T}^{2}{\bf r}^2$ with harmonic length $d = \sqrt{\hbar/(M\omega)}\gg 2\pi/q$, the wavelength of the stripe, we can apply Thomas-Fermi approximation, and the condensate wavefunction is given by Eq.(\ref{op}), (\ref{A2}) and (\ref{apm}) with chemical potential $\mu$ in Eq.(\ref{A2}) replaced by $\mu({\bf r})= \mu-V({\bf r})$, i.e. 
for Phase {\bf (I)}, 
\begin{eqnarray}
\Phi_{m} &=  \sqrt{\frac{\mu({\bf r})- E_{o}}{G_{o}} } [   a^{o}_{+}  e^{ip_{+}x} \left( \begin{array}{c} i {\rm sin}\frac{\theta}{2} \\ {\rm cos}\frac{\theta}{2}\end{array} \right)  \nonumber \\ 
&  +   e^{i\gamma}  a^{o}_{-} e^{ip_{-}x} 
\left( \begin{array}{c} i {\rm cos}\frac{\theta}{2} \\ {\rm sin}\frac{\theta}{2} \end{array} \right)]
\end{eqnarray}
The density profile $n_{1}({\bf r})$  for the $m=1$ spin component along $\hat{\bf y}$ is shown in Figure 4 for the parameters $\theta = \frac{1}{4}\pi$ $N=2.5\times 10^5$ etc.  

Apart from the stripe structure, the presence of these phases can be
detected by measuring the displacement of the atomic cloud in expansion after the trap is turned off.  For single dressed state with 
momentum $p_{+}$ and $p_{-}$, the atom cloud will displace in  $x$-direction by an amount determined by $p_{+}$ or $p_{-}$.  
For the condensate with two dressed states,  the system after expansion will separate into two atom clouds moving with different momenta. 

\begin{figure}[h]
\includegraphics[width=0.4 \textwidth]{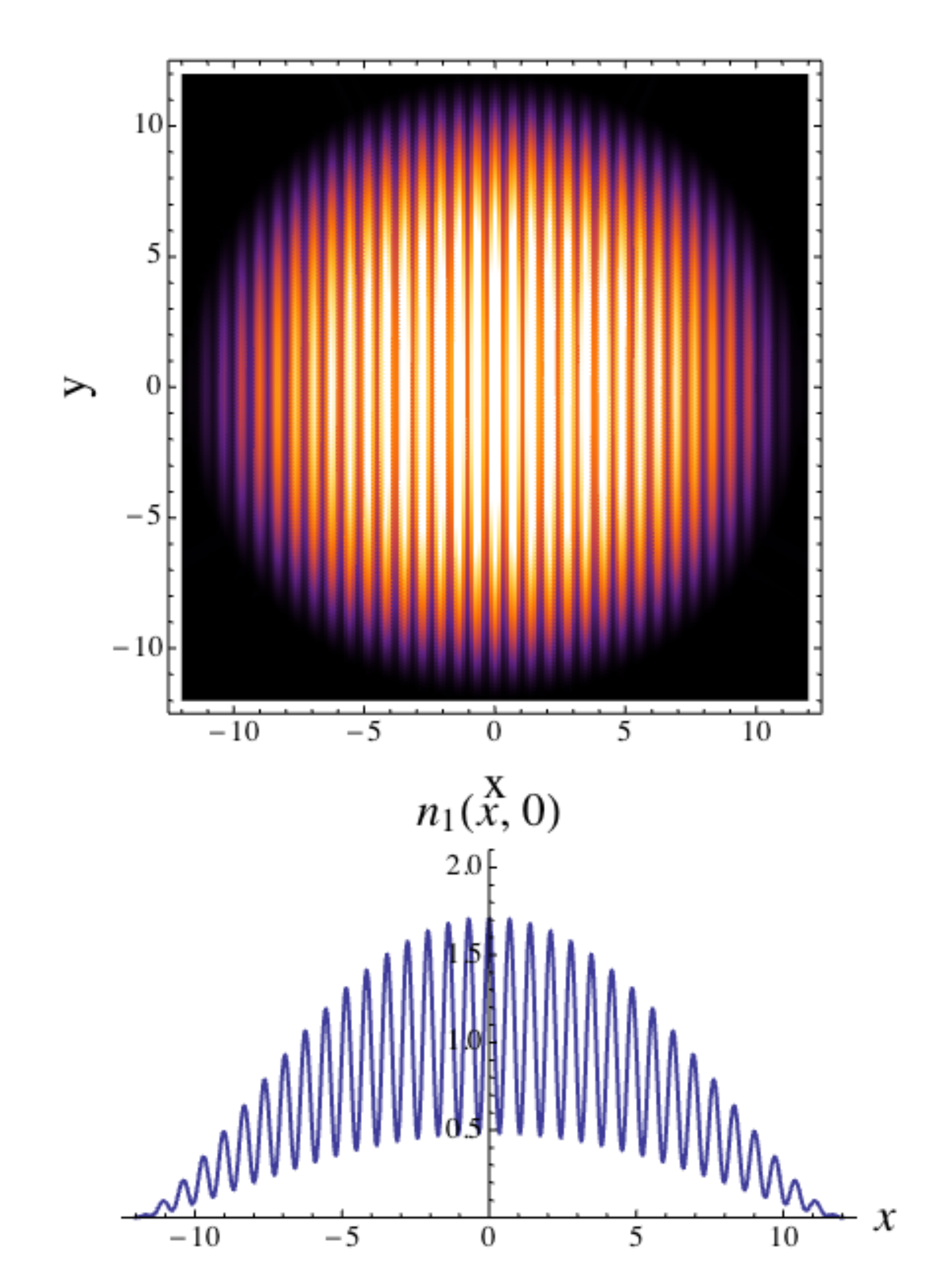}
\caption{The upper figure is the column density $\widetilde{n}_1(x,y) = \int {\rm d}z n_1(x,y,z)$. The lower frame is
$\widetilde{n}_1(x,0)$. The period of oscillation is $\pi/k_{o}$. The contrast of oscillation at the center  is 70$\%$. 
Our calculation is performed for $^{87}$Rb with $N=2.5\times 10^5$ atoms, $q=1.56\times 10^7 m^{-1}$, $\theta =\frac{1}{4} \pi$, $\hbar\Omega_{R}=h\times 7.1$KHz, cloud size $R_{TF}=20\mu$m\cite{ian1}. The values $\alpha$ and $\beta$ used are given by $\alpha=\frac{1}{4}\alpha_c$ and $\beta=\frac{1}{4}\beta_c$. The 
length displayed is  in units of the laser wavelength $804.3 nm$ in ref.\cite{ian1} .}  
\label{stripe}
\end{figure}


Our discussions here focus on current experiments. It is, however, useful to put them in a more general context.
Let us return to the single particle hamiltonian $H_{mn}= \delta_{mn}{\bf p}^2/2M + W_{mn}({\bf r})$, where $W({\bf r})$ is a matrix in a spin-$F$ space, $m= -F, -F+1, \cdots, F$.  Let $|n\rangle_{\bf r}$ be the eigenstate of $W$ with energy ${\cal E}_{n}$, which we label as 
${\cal E}_{F}< {\cal E}_{F-1}< ... <{\cal E}_{-F} $ in ascending order. The unitary matrix  $U({\bf r})$ that diagonalizes $W$ in the original spin basis is $U_{mn}({\bf r}) = \langle m| n\rangle_{\bf r}$, and 
$[U^{\dagger}({\bf r}) W({\bf r}) U({\bf r})]_{mn}= {\cal E}_{m}\delta _{mn}$. In the frame where $W$ is diagonal, the hamiltonian is 
$[U^{}({\bf r}) H U^{\dagger}({\bf r})]_{mn}= \frac{1}{2M}[({\bf p} + {\bf A})^2]_{mn} + {\cal E}_{n}\delta_{mn}$, where 
the ${\bf A}_{mn}= \langle m|_{\bf r}{\bf p} |n\rangle_{\bf r} $ is the gauge field emerging in this frame \cite{Zee}.  The gauge field is non-abelian if $[A^{i}, A^{j}]\neq 0$.
For the NIST experiment, one  restricts to the lowest two states of spin-1 system. When the field gradient $G\neq 0$, ${\bf A}$ has both $x$ and $y$ component and is non-abelian. 
The $G=0$ case that we considered is special, since ${\bf A}$ has only one component.  However, this uniform gauge field can not gauged away without changing the hamiltonian due to the presence of the Rabi term  $\Omega_{R}\tau_{1}$ in Eq.(\ref{h}).
The phenomena associated with the non-commutativity ${\bf A}$ caused by field gradients are sufficiently rich that they deserve  separate discussions. 


T.L. Ho would like thank Ian Spielman for discussions of his experiments. 
This work is supported by NSF Grant DMR-0907366 and by DARPA under the Army Research Office Grant Nos. W911NF-07-1-0464, W911NF0710576.

\end{document}